\documentclass[preprint,aps,showpacs]{revtex4}

\usepackage{graphicx}
\usepackage{bm}       
\usepackage{mathptmx} 

\begin{document}

\title{Flavor extrapolations and staggered fermions}
\author{Michael Creutz}
\affiliation{
Physics Department, Brookhaven National Laboratory\\
Upton, NY 11973, USA
}

\begin{abstract}
A popular approximation in lattice gauge theory is an extrapolation in
the number of fermion species away from the four fold degeneracy
natural with the staggered fermions formulation.  I show that at
finite lattice spacing and for an odd number of flavors this
extrapolation misses terms which, on general principles, must be
present in the continuum theory.  For a correct continuum limit, this
forces unphysical singularities in parameter regions where continuum
physics is smooth and all physical particles are massive.  These
singularities are not expected with other lattice regulators.
Finally, I argue that unnatural constraints on certain correlation
functions appear even when all quarks are massive.
\end{abstract}
\maketitle

Lattice gauge theory provides a powerful tool for the investigation of
non-perturbative phenomena in strongly coupled field theories, such as
the quark confining dynamics of the strong interactions.  However
numerical calculations are quite computer intensive, strongly
motivating approximations that reduce this need.  One such, the
valence or quenched approximation
\cite{Weingarten:1980hx,Hamber:1981zn}, introduces rather uncontrolled
uncertainties, but with the growth in computer power, its use is
currently being eliminated.

Another popular approximation \cite{Aubin:2004ej,Aubin:2004fs} arises
from the simplicity of the staggered fermion formulation \cite{
Kogut:1974ag,Susskind:1976jm,Sharatchandra:1981si}.  With only one
Dirac component on each site, the large matrix inversions involved
with conventional algorithms are substantially faster than with other
fermion formulations.  However the approach and its generalizations
are based on a discretization method that inherently requires a
multiple of four fundamental fermions.  The reasons for this are
related to the cancellation of chiral anomalies.  To apply the
technique to the physical situation of two light and one intermediate
mass quark requires an extrapolation down in the number of fermions.
As usually implemented, the approach involves taking a root of the
fermion determinant inside standard hybrid Monte Carlo simulation
algorithms.  This step has not been justified theoretically.  The
purpose of this note is to show that at finite lattice spacing this
reduction inherently misses certain required terms in the chiral
expansion for the continuum theory.  To reproduce these terms in the
continuum limit requires the introduction of unphysical singularities
which persist at finite volume and in regions of parameter space where
there are no physical massless particles.  Even in regions where the
physical fermion determinant is positive definite, the procedure
imposes unexpected and non-trivial constraints on correlations between
certain fermion bilinears.

The method has its roots in the ``naive'' discretization of the
derivatives in the Dirac equation
\begin{equation}
\overline\psi \gamma_\mu \partial_\mu \psi \rightarrow
{1\over 2 a} \overline\psi_x \gamma_\mu (\psi_{x+ae_\mu}-\psi_{x-ae_\mu})
\end{equation}
with $a$ denoting the lattice spacing.  Fourier transforming to
momentum space, the momentum becomes a trigonometric function
\begin{equation}p_\mu\rightarrow {1\over i a}
(e^{iap_\mu}-e^{-iap_\mu})={1\over a}\sin(ap_\mu)
\end{equation}
The natural range of momentum is $-\pi/a <p_\mu \le \pi/a$.  The
doubling issue is that the propagator has poles not just at small
momentum, but also when any component is near $\pi$ in magnitude.
These all contribute as intermediate states in Feynman diagrams; so,
the theory effectively has $2^4=16$ fermions.  I refer to these
multiple states as ``doublers'' or ``flavors'' in the following
discussion.

Note that the slope of the sine function at $\pi$ is opposite to that
at 0.  This can be absorbed by changing the sign of the corresponding
gamma matrix.  This changes the sign of $\gamma_5$ as well; so, the
doublers divide into different chirality subsets.  The determinant of
the Dirac operator is not simply the sixteenth power of a single
determinant.

Without a mass, the naive action has an exact chiral symmetry of the
kinetic term under
\begin{eqnarray}
&\psi\rightarrow e^{i\theta\gamma_5}\psi\cr
&\overline\psi\rightarrow \overline\psi e^{i\theta\gamma_5}
\label{chiral}
\end{eqnarray}
The conventional mass term is not invariant under this rotation
\begin{equation}
m\overline\psi \psi\rightarrow m\overline\psi \psi \cos(2\theta)
+im\overline\psi\gamma_5 \psi\sin(2\theta)
\end{equation} 
Thus any mass term of the form on the right hand side of this relation
can have theta rotated away.  This is consistent with known anomalies
since this is in reality a flavor non-singlet chiral rotation.  The
different species use different signs for $\gamma_5$.  As special
cases, in this theory $m,-m,$ and $\pm i\gamma_5 m$ are all physically
equivalent.

To arrive at the staggered formulation, note that whenever a fermion
hops between neighboring sites in direction $\mu$, it picks up a
factor of $\gamma_\mu$.  An arbitrary closed fermion loop on a
hypercubic lattice gives a product of many gamma factors, but any
particular component always appears an even number of times.  Bringing
them through each other using anti-commutation, the net factor for any
loop is proportional to unity.  Gauge fields don't change this fact
since they just involve $SU(3)$ phases on the links.  So if a fermion
starts in one spinor component, it returns to the same component after
the loop.  The 4 Dirac components give 4 independent theories.  There
is an exact $SU(4)$ symmetry.  Without a mass term, this is actually
an exact $SU(4)\otimes SU(4)$ chiral symmetry \cite{Karsten:1980wd}.

Staggered fermions single out one component on each site.  Which
component depends on the gamma factors to get to the site in question
from one starting site.  Ignoring the other components reduces the
degeneracy from 16 to 4.  The process brings in various oscillating
phases from the gamma matrix components.  One explicit projection that
accomplishes this is (using integer coordinates and the convention
$\gamma_5=-\gamma_1\gamma_2\gamma_3\gamma_4$ with Euclidean gamma
matrices)
\begin{equation}
P=P^2={1\over 4} \left(1+i\gamma_1\gamma_2 (-1)^{x_1+x_2}
+i\gamma_3\gamma_4
(-1)^{x_3+x_4}+\gamma_5(-1)^{x_1+x_2+x_3+x_4}\right)
\label{projection}
\end{equation} 
Note that some degeneracy must remain.  No chiral
breaking appears in the action, and all infinities are removed.  Thus
there is no way for the anomaly to appear.  It is canceled between the
remaining species.  Furthermore, the naive replacement
$\psi\rightarrow\gamma_5\psi$ exactly relates the theory with mass $m$
and mass $-m$.  With 4 flavors this symmetry is allowed since there is
a flavored chiral rotation that gives it.  The doublers still are in
chiral pairs.

To proceed I sketch how a typical simulation with fermions proceeds.
For a generic fermion matrix $D$, the goal of a the simulation is to
generate configurations of gauge fields $A$ with a probability
\begin{equation}
P(A)\propto \exp(-S_g(A)+N_f {\rm Tr}\ \log(D(A)))
\end{equation}
Here $S_g$ is the pure gauge part of the action.  With some algorithms
additional commuting ``pseudo-fermion'' fields are introduced
\cite{Fucito:1980fh,Scalapino:1981qs}, but these details are not
important to the following discussion.  With staggered or naive
fermions the eigenvalues of $D$ all appear in complex conjugate pairs;
thus, the determinant is non-negative as necessary for a probability
density.

In hybrid Monte Carlo schemes \cite{Duane:1987de} auxiliary
``momentum'' variables $P$ are introduced, one for each degree of
freedom in $A$.  The above distribution is generalized into
\begin{equation}
P(A,P)\propto \exp\left(-S_g(A)+N_f {\rm Tr}\ \log(D(A))+\sum
P_i^2/2\right)
\label{pap}
\end{equation} 
As the momenta are Gaussian random variables, it is easy to generate a
new set at any time.  For the gauge fields one sets up a
``trajectory'' in a fictitious ``Monte Carlo'' time variable $\tau$
and uses the exponent in (\ref{pap}) as a classical Hamiltonian
\begin{equation}
H=\sum P_i^2/2+V(A)
\end{equation}
with the ``potential''
\begin{equation}
V(A)=-S_g(A)+N_f {\rm Tr}\log(D(A)).
\end{equation}
The Hamiltonian dynamics
\begin{eqnarray}
&{dA_i\over d\tau}=P_i\cr
&{dP_i\over d\tau}= F_i(A)=-{\partial V(A)\over \partial A}
\end{eqnarray}
conserves energy and phase space.  Under such evolution the
equilibrium ensemble stays in equilibrium, a sufficient condition for
a valid Monte Carlo algorithm.  After evolution along a trajectory of
some length $\tau$, discretized time steps $\delta\tau$ can introduce
finite step errors and give a small change in the ``energy.''  The
hybrid Monte Carlo algorithm corrects for this with a Metropolis
accept/reject step on the entire the trajectory.  The trajectory
length and step size are parameters to be adjusted for reasonable
acceptance.  After the trajectory one can refresh the momenta by
generating a new set of gaussianly distributed random numbers.  The
procedure requires the ``force'' term
\begin{equation}
 F_i(A)=-{\partial V(A)\over \partial A}
={\partial S_g(A)\over \partial A}-N_f{\rm Tr} 
\left( D^{-1} {\partial D(A)\over \partial A}\right).
\label{force}
\end{equation}
To calculate the second term requires an inversion of the sparse
matrix $D$ applied to a fixed vector.  Standard linear algebra
techniques such as a conjugate gradient algorithm can accomplish this.
In practice this step is the most time consuming part of the
algorithm.

Returning to staggered fermions, one would like to eliminate the
unwanted degeneracy by a factor of four.  One attempt to do this
reduction involves an extrapolation in the number of flavors.  In the
molecular dynamics trajectories for the simulation of the gauge field,
the coefficient of the fermionic force term in Eq.~(\ref{force}) is
arbitrarily reduced from $N_f$ to $N_f/4$, where $N_f$ is the desired
number of physical flavors.  Although not proven, this seems
reasonable when $N_f$ is itself a multiple of four.  The controversy
arises for other values of $N_f$.

Here I argue that the procedure is an approximation that incorrectly
predicts certain qualitative behaviors.  The issue is clearest in the
chiral limit when when $N_f$ is odd.  For the staggered theory, the
fermion determinant is a function of $m^2$.  The surviving chiral
symmetry gives equivalent physics for either $m$ or $-m$.  The primary
problem with the extrapolation appears at this point.  It is well
known that with an odd number of flavors, physics has no symmetry
under changing the sign of the mass \cite{DiVecchia:1980ve,
Creutz:1995wf, Creutz:2003xu}.  The most dramatic demonstration of
this appears in the one flavor theory.  In this case anomalies break
all chiral symmetries and no Goldstone bosons are expected.  The
theory behaves smoothly as the mass parameter passes through zero.
The lightest meson, call it the $\eta^\prime$, acquires a mass through
anomaly effects, and the lowest order quark mass corrections are
linear
\begin{equation}
m_{\eta^\prime}^2(m)=m_{\eta^\prime}^2(0)+ c m
\label{etaprime}
\end{equation} 
Such a linear dependence in a physical observable is immediately
inconsistent with $m\leftrightarrow -m$ symmetry.

The one flavor case is perhaps a bit special, but there are similar
problems with the three flavor situation \cite{Creutz:2003xu}.
Identify the quark bi-linear with an effective chiral field
$\overline\psi_a\psi_b\sim \Sigma_{ab}.$ Here $a$ and $b$ are flavor
indices.  The $SU(3)\otimes SU(3)$ chiral symmetry of the massless
theory is embodied in the transformation
\begin{equation}
\Sigma \rightarrow g_L^\dagger \Sigma g_R
\end{equation}
with $g_L,g_R \in SU(3)$.  For positive mass, $\Sigma$ should have an
expectation value proportional to the $SU(3)$ identity $I$.  This is
not equivalent to the negative mass theory because $-I$ is not in
SU(3).  Indeed, for negative mass it is expected that the infinite
volume theory spontaneously breaks $CP$ symmetry, with
$\langle\Sigma\rangle\propto e^{\pm 2\pi i/3}$
\cite{Montvay:1999gn,Creutz:2003xu}.

These qualitative effective Lagrangian arguments are quite powerful
and general.  Another way to see the one flavor behavior is to start
with a larger number of flavors, say 3 or 4, and make the masses
non-degenerate.  As only one of the masses passes through zero, the
behavior for the lightest meson mimics that in Eq.~(\ref{etaprime}).
Extrapolated staggered quarks with their symmetry under taking any
quark mass to its negative will miss the linear term.

Note that with degenerate quarks these arguments become sharpest at
finite volume.  In the infinite volume limit the multiflavor massless
theory exhibits spontaneous chiral symmetry breaking and a
non-analytic behavior in the mass at $m=0$.  But at finite volume and
with a finite lattice spacing all physical quantities being considered
are analytic.  The only way the extrapolation from $N_f$ to $N_f/4$ to
give correct physics at finite volume would be for it to introduce
unphysical nonanalytic terms.

Small real eigenvalues of the Dirac operator are responsible for these
effects.  The odd terms come from topological structures in the gauge
fields \cite{'tHooft:fv}.  For small mass in the traditional continuum
discussion, $|D|\sim m^\nu$ with $\nu$ the winding number of the gauge
field.  The condensate
\begin{equation}
\langle\overline\psi \psi\rangle=
{1\over Z}\int (dA) |D|^{N_f} e^{-S_g(A)}\ {\rm Tr}\ D^{-1}
\end{equation}
receives a contribution going as $m^{N_f-1}$ from the $\nu=1$ sector.
For the one flavor case, this is an additive constant.  This constant
will be missing from the extrapolated staggered theory because of the
symmetry in Eq.~(\ref{chiral}).  This phenomenon is also responsible
for the fact that a single massless quark is not a well defined
concept \cite{Creutz:2004fi}.

For the general odd flavor case, the odd winding number terms have the
opposite symmetry under the sign of the mass than the even terms,
although with more flavors this starts at a higher order in the mass.
For 3 flavors the condensate at finite volume will display a $m^2$
correction to the leading linear behavior.  The extrapolation down
from the staggered 4 flavor theory will not see this.  While the zero
modes of the Dirac operator are suppressed at finite volume, they do
not vanish.

This mechanism emphasizes an important distinction between staggered
and other fermion formulations.  With staggered fermions there is no
exact index relation between the zero modes of the Dirac operator and
the topology of the gauge fields \cite{Smit:1986fn}.  Isolated real
eigenvalues of the Dirac operator are a robust concept for many
formulations, such as Wilson \cite{Wilson:1975id,Creutz:2005rd},
domain wall \cite{Furman:1994ky}, and overlap \cite{Neuberger:1997fp}
fermions.  However this is not expected for the staggered fermions,
where the real part of all eigenvalues is pinned to the mass.  As
discussed above, the exact chiral symmetry is actually a flavored
chiral rotation.  The respective species transform with different
signs for $\gamma_5$.  Because of this, the corresponding zero modes
generically can mix.  Small variations in the gauge field can split
the degenerate real eigenvalues apart into the complex plane.  Unlike
with the overlap, those gauge configurations where the staggered
matrix has exactly real eigenvalues is expected to be a set of measure
zero.  For all other configurations the determinant is non-vanishing
and analytic in $m^2$ around $0$.

While I have shown diseases with the chiral behavior of extrapolated
staggered fermions at finite cutoff, technically I have not proven
that these problems survive as the cutoff is removed
\cite{Durr:2004ta,Bernard:2006vv,Durr:2006ze}.  Indeed, in field
theory we are accustomed to the non-commutation of certain limits, such as
vanishing mass and infinite volume when a symmetry is being
spontaneously broken.  In that case the mass and the volume are both
infrared issues.  As the lattice is an ultraviolet regulator and the
chiral issues raised here involve long distance physics, it seems
peculiar for the order of these limits to affect each other.
Nevertheless, suppose that taking the cutoff to zero before taking the
massless limit does give the correct physics.  Then the regulator must
introduce singularities that are not present in the continuum theory.

The issue is again clearest for the one flavor theory, where in the
continuum the condensate, $\langle \overline\psi\psi \rangle$
appropriately renormalized, does not vanish and is smoothly behaved
around m=0.  Analyticity in the mass is expected with a radius of
order the eta-prime mass-squared over the typical scale of the strong
interactions, $\Lambda_{\rm qcd}$.  Now turn on the extrapolated
staggered regulator.  At $m=0$, $\langle \overline\psi\psi \rangle$
must suddenly jump to zero.  For every eigenvalue of the staggered
fermion matrix at vanishing mass, its negative is also an eigenvalue.
Thus configuration by configuration the trace of $D^{-1}$, and thus
the condensate, is identically zero.  Furthermore, due to confinement
and the chiral anomaly, this unphysical jump occurs both at finite
volume and in the absence of any massless physical particles for the
continuum theory.

This issue generalizes to the multiflavor theory with non-degenerate
quark masses.  The proposed regulator forces the condensate associated
with any given species to vanish with the corresponding mass, in
contradiction with the continuum behavior expected from effective
Lagrangian analysis.  Physical observables at specific points in
parameter space where continuum physics is smooth are forced to
develop infinite derivatives with respect to the cutoff as it is
removed.  Even if this occurs only in the vicinity of isolated points,
this seems an absurd behavior for an ultraviolet regulator and is in
strong contrast to more sensible schemes such as Wilson fermions
\cite{Wilson:1975id}.

Despite this highly unphysical behavior, certain authors
\cite{Bernard:2006vv} continue to advocate that, while ugly, the
continuum limit could be correct as long as one avoids these
singularities.  This, however, requires some rather peculiar relations
amongst correlation functions even for quark masses in regimes where
the fermion determinant is expected to be positive definite.  Consider
the case of two flavor QCD with quark masses $m_u$ and $m_d$.
Complexifying the mass terms in the usual way
\begin{equation}
\sum_{a=u,d}
{\rm Re}\ m_a\ \overline\psi^a\psi^a
 +i\ {\rm Im}\ m_a\ \overline\psi^a\gamma_5\psi^a
\end{equation}
the physical theory is invariant under the flavored chiral rotation
\begin{eqnarray}
&m_u\rightarrow e^{i\theta}m_u\cr
&m_d\rightarrow e^{-i\theta}m_d
\end{eqnarray}
Due to the chiral anomaly, it must not be invariant under the singlet
chiral rotation
\begin{eqnarray}
&m_u\rightarrow e^{i\theta}m_u\cr
&m_d\rightarrow e^{i\theta}m_d
\end{eqnarray}
The symmetry in mass parameter space requires that the rotations of
the up and down quark masses be in opposite directions.

Now formulate this theory with two independent staggered fermions, one
for the up and one for the down quark, each reduced using the rooting
procedure.  From Eq.~(\ref{projection}), the corresponding
complexification of the staggered mass term takes the form
\begin{equation}
\sum_{a=u,d} ({\rm Re}\ m_a + i S(j)\ {\rm Im}\ m_a)\ \psi^\dagger(j)\psi(j)
\end{equation}
with $S(j)$ being $\pm 1$ depending on the parity of the site $j$.
The issue arises from the fact that that the staggered fermion
determinant, and therefore the path integral, are exactly invariant
under $m\rightarrow e^{i\theta}m$ for either the up or the down quark.
This is too much symmetry in parameter space.  It is precisely this
extra symmetry that forces the unphysical singularities mentioned
above.  But the consequences extend to positive masses as
well.  Considering an infinitesmal rotation on the up quark alone, we
have
\begin{equation}
\left .{d Z \over d \theta_u}\right\vert_{\theta_u=0}=0
\end{equation}
This means that correlators of 
\begin{equation}
\sum_j S(j)\psi_u^\dagger(j) \psi_u(j)
\end{equation}
with any operators not involving the up quark field must vanish
identically.  This occurs configuration by configuration and for any
quark masses.  As one example,
\begin{equation}
\left\langle
\psi_d^\dagger(k)\psi_d(k)\
 \sum_j S(j)\psi_u^\dagger(j) \psi_u(j)\right\rangle=0
\end{equation}
requires a delicate cancellation of the expected contribution from the
$\pi_0$ at large distances against short distance physics.  As the
former contribution diverges as the quark mass goes to zero, this
cancellation seems highly contrived and is unexpected in other
formulations.  Note that for the unextrapolated theory the
cancellation occurs naturally between the additional bosons of the 8
flavors.  But the two flavor theory should only have one neutral pion.

I have argued that the extrapolation involved in extrapolating the
staggered quark formulation of lattice gauge theory to physical
numbers of species is unlikely to become exact in the continuum limit.
The behavior in the chiral limit is incorrect at finite lattice
spacing, forcing unphysical singularities.  For all mass values,
including where the physical fermion determinant is positive definite,
certain non-trivial correlations are unexpectedly forced to vanish.

The approximation may still be reasonable for some observables, most
particularly those involving only flavor non-singlet particles.  But
any predictions for which anomalies are important are particularly
suspect.  This includes the $\eta^\prime$ mass, but also more mundane
quantities such as the lightest baryon mass, which in the chiral limit
also receives a non-perturbative contribution.

\section*{Acknowledgments}
This manuscript has been authored under contract number
DE-AC02-98CH10886 with the U.S.~Department of Energy.  Accordingly,
the U.S. Government retains a non-exclusive, royalty-free license to
publish or reproduce the published form of this contribution, or allow
others to do so, for U.S.~Government purposes.

\end{document}